# A Comparison between Network-Controlled Repeaters and Reconfigurable Intelligent Surfaces

Hao Guo*, Charitha Madapatha*, Behrooz Makki‡, Boris Dortschy‡, Lei Bao‡, Magnus Åström‡, and Tommy Svensson*
*Department of Electrical Engineering, Chalmers University of Technology, Gothenburg, Sweden
‡Ericsson Research, Ericsson AB, Sweden
Email: {hao.guo, charitha, tommy.svensson}@chalmers.se, {behrooz.makki, boris.dortschy, lei.bao, magnus.astrom}@ericsson.com

*Abstract*—Network-controlled repeater (NCR) has been recently considered as a study-item in 3GPP Release 18, and the discussions are continuing in a work-item. In this paper, we introduce the concept of NCRs, as a possible low-complexity device to support for network densification and compare the performance of the NCRs with those achieved by reconfigurable intelligent surfaces (RISs). The results are presented for the cases with different beamforming methods and hardware impairment models of the RIS. Moreover, we introduce the objectives of the 3GPP Release 18 NCR work-item and study the effect of different parameters on the performance of NCR-assisted networks. As we show, with a proper deployment, the presence of NCRs/RISs can improve the network performance considerably.

## I. INTRODUCTION

To improve the coverage and support the increasing number of user equipments (UEs), different methods are considered, among which network densification and millimeter wave (mmw) communications are considered important contributions. Network densification refers to the deployment of multiple access points of different types in, e.g., metropolitan areas. Particularly, it is expected that in future small nodes, such as relays, integrated access and backhaul (IAB), repeaters, etc., may be deployed to assist the communications [1].

During the 3GPP Release 16 and 17, IAB was specified as the main relaying technique in 5G NR [2]-[5], and the discussions are continuing in Release 18 to introduce mobility for IAB nodes [6]-[8]. Considering decode-and-forward relaying setup, the IAB network well extends the coverage and/or increase the throughput [9]-[10]. However, while the coverage of the IAB is substantially larger than the coverage of, e.g., Release 17 and 18 repeaters, it is considered a relatively complex node and thereby, depending on the deployment, we may require simpler nodes with low complexity for, e.g., blind spot removal. Here, radio frequency (RF) repeater [11] is one candidate type of node that simply amplifies-and-forwards every signal that it receives. However, RF repeaters lack in, e.g., accurate beamforming which may limit its efficiency at high frequencies. Moreover, as we explain in the following, adding control of repeater operation by a base station (BS) gives the chance to improve the data transmission and/or the energy efficiency of the repeaters.

With this background, a study-item was performed in 3GPP Release 18, finalized in August 2022, in which the potentials and the challenges of network-controlled repeaters (NCRs) have been evaluated. The 3GPP discussions on NCR are currently continuing in a work-item. In simple words, NCR is a normal repeater with beamforming capabilities and under the network control.

From a different perspective, reconfigurable intelligent surface (RIS) may be a possible technology for beyond 5G networks (RIS has not yet been considered in the 3GPP discussions). It enables to control the electromagnetic properties of the RF waves by performing an intelligent adaptation of the phase shift towards the desired direction [12]–[16]. In general, intelligent surfaces are electromagnetically active artificial structures with beamforming capabilities that can be used to reshape the propagation environment such as to improve capacity, coverage and energy efficiency. Both NCR and RIS emit incoming signals without decoding. There are still ambiguities about the detailed differences of the NCRs and the RISs. A simple explanation is that an RIS is an NCR with negative gain. In general, an RIS is expected to be a simpler node with less focused beamforming capability/accuracy and without active amplification. That is, RIS may be capable of signal reflection via adapting a phase matrix, while the NCR is capable of advanced beamforming with power amplification. As opposed to the NCR, RIS does not amplify the signal but also not the noise. Moreover, an RIS has only one beamforming matrix, whereas an NCR can do separate beamforming at receiver and transmitter side. This should be an advantage for the NCR when controlling the interference in, e.g., multi-user scenarios. In this way, it is interesting to compare the performance of the RISs and the NCRs.

In this paper, we introduce the concept of NCRs, as considered by the 3GPP Release 18, and compare its performance with those achieved by RISs. We present the objectives and the open research issues to be discussed in 3GPP Release 18. Moreover, with different hardware impairment (HWI) models of the RIS and beamforming methods, we present illustrative simulation results on NCRs, to clarify its concept, and compare the performance of NCR-assisted networks with those achieved in RIS-assisted networks. As we show, with a proper deployment, NCR- and RIS-assisted networks improve the network performance considerably. Moreover, for a broad range of parameter settings, the NCR-assisted network outperforms the RIS-assisted networks, in terms of UEs' achievable rate.

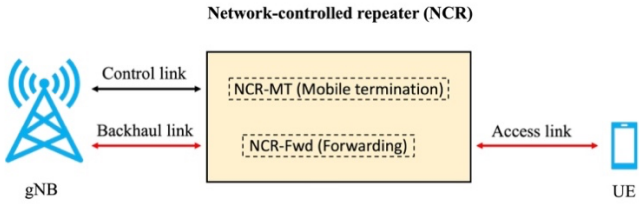

Fig. 1. Schematic of an NCR-based system.

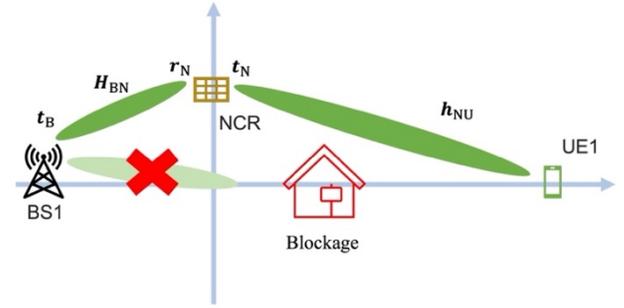

Fig. 2. NCR-assisted MISO DL system with a single link.

## II. Network-controlled repeater

In May 2022, 3GPP started a new study-item on NCRs. The study-item was finalized in August 2022, and 3GPP has recently started the NCR work-item. In simple words, NCR is a repeater with beamforming capabilities that can receive and process side control information from the network. Such side control information could allow an NCR to perform its amplify-and-forward operation in an efficient manner. In general, the NCR can be considered as a network-controlled "*beam bender*" relative to the gNB. In this way, NCR is logically part of the gNB for all management purposes, i.e., it can be assumed that the NCR is deployed and under the control of the operator. NCR is based on amplify-and-forward relaying scheme, and in 3GPP it is limited to single-hop communication in stationary deployments. Particularly, the NR work-item description [17] makes the following assumptions for the NCRs:

- NCRs are in-band RF repeaters used for extension of network coverage on FR1 and FR2 bands[1] based on the NCR model in TR38.867 [18].
- For only single hop stationary NCRs.
- The NCR is transparent to the UE.
- NCR can maintain the gNB-repeater link and repeater-UE link simultaneously.

Fig. 1 shows the schematic of an NCR [17]. The NCR-mobile termination (MT) is defined as a functional entity to communicate with a gNB via Control link (C-link) to enable the information exchanges, e.g., side-control information. The C-link is based on NR Uu interface. The NCR-forward (Fwd) is defined as a functional entity to perform the amplify-and-forwarding of uplink/downlink (UL/DL) RF signal between the gNB and the UE via backhaul link and access link (see Fig. 1 for the illustration of different links). The behavior of the NCR-Fwd will be controlled according to the side control information received by the NCR-MT from the controlling gNB.

In general, the NCR-MT and the NCR-Fwd modules could be operating at the same, overlapping, or different frequencies. However, controlling the backhaul link, i.e., the link between the gNB and the NCR-Fwd, will be much simplified if the NCR-MT and NCR-Fwd operate within the same band. Particularly, at least one of the NCR-MT's carrier(s) should be within the set of carriers forwarded by the NCR-Fwd in the same frequency band.

The NCR is equipped with an antenna configuration, where a signal is first received in DL (or UL), and, after power amplification and with proper beamforming, is forwarded in DL (or UL). Since the NCR-Fwd module only amplifies and (analogously) beamforms the signal, no advanced digital receiver or transmitter chains may be required. The NCR-MT module is used to exchange control and status signaling with the gNB that is controlling the NCR, through the C-link as shown in Fig. 1. For this, the NCR-MT module supports at least a sub-set of UE functions. As baseline, the same large-scale properties of the channel are expected to be experienced by C-link and backhaul link (at least when the NCR-MT and NCR-Fwd operating in same carrier). That is, as baseline, the same transmission configuration indicator (TCI) states as C-link are assumed for beam at NCR-Fwd for backhaul link if the NCR-MT's carrier(s) is within the set of carriers forwarded by the NCR-Fwd. For the transmission/reception of C-link and backhaul link by NCR:

- The DL of C-link and DL of backhaul link can be performed simultaneously or TDM:ed;
- The UL of C-link and UL of backhaul link can be performed at least TDM:ed.

The simultaneous transmission of the UL of C-link and UL of backhaul link is subject to NCR capability (see [19] and [20] for the details of the specifications agreed in 3GPP RAN WG1 meetings #109e and #110, respectively).

In the following, the indices b, n, r, and u stand for the BS, the NCR, the RIS and the UE, respectively. To further illustrate the concept of the NCR, consider the network model presented in Fig. 2. We consider an NCR-assisted multiple-input-single-output (MISO) DL system, where the BS is equipped with $N_b$ transmit antennas transmitting to a single-antenna UE. To improve the system performance in the presence of, e.g., blockage, one NCR with $N_n$ antennas on both the BS- and access-side of the NCR is deployed. We ignore the direct BS-UE link, due to, e.g., blockage. In this way, the power of the useful signal received at the UE can be expressed as

$$S = Pg\gamma_{bn}\gamma_{nu}, \quad (1)$$

where $P$ is the transmit power at the BS and $g$ is the amplification gain applied by the NCR. Also, in such a DL system,

$$\gamma_{i,j} = \boldsymbol{r}_j \boldsymbol{H}_{i,j} \boldsymbol{t}_i \quad (2)$$

represents the combined effective channel from node $i$ (the BS or the NCR depending on the considered link) to node $j$ (the NCR or the UE depending on the considered link) with transmit and receive beamforming, i.e., $\boldsymbol{t}_i$ and $\boldsymbol{r}_j$, respectively. Also, $\boldsymbol{H}_{i,j}$ denotes the channel between transmitter $i = \{b, n\}$, and receiver $j = \{n, u\}$. Assuming perfect channel state information (CSI), the transmit and receive beamformer can be obtained by, e.g., maximum-ratio transmission. The total noise power received by the UE can be calculated as

$$NP = \sigma_U^2 + \sigma_N^2 g \gamma_{nu}. \quad (3)$$

where $\sigma_N^2$ and $\sigma_U^2$ are the variance of the AWGN at the NCR and UE, respectively. In this way, the received signal-to-noise ratio (SNR) at the UE is

$$\text{SNR} = \frac{Pg\gamma_{bn}\gamma_{nu}}{\sigma_U^2 + \sigma_N^2 g \gamma_{nu}}. \quad (4)$$

---

[1] FR1 (FR: frequency range) defines bands in the sub-6 GHz spectrum (although 7125 MHz is the maximum) and FR2 defines bands in the mmw spectrum.

Also, the achievable rate is given by

$$R = Q\log_2(1 + \text{SNR}), \quad (5)$$

with $Q$ being the bandwidth. Note that, while we consider a fixed amplification gain for the NCR, it may also have a maximum output power constraint, i.e., the NCR output power cannot exceed a certain threshold $\hat{P}$. In this way, while (1)-(3) present the signal model for the cases with a fixed amplification gain, in general the output power of the NCR is determined as $P_n = \min\left(\hat{P}, g(\sigma^2 + P\gamma_{bn})\right)$. Note that, motivated by the fact that the NCR may have a better hardware compared to the RIS, we consider the HWI only for the RISs. Studying the effect of HWIs, self-interference, etc. on the performance of NCRs is an interesting extension of the paper.

### III. RIS-ASSISTED COMMUNICATIONS

It is interesting to compare the performance of the NCRs with the RISs as they are both based on the same concept of forwarding the signal without decoding. Similar to NCR, RIS may be helpful by providing alternative paths when the direct link from the BS to the UE faces deep fading and/or blockage. Moreover, with a proper deployment, RIS can cover different areas with different service requirements, which makes it possible to bypass certain blockages.

For the $M$-element RIS-assisted link, the received signal for the UE can be expressed as

$$y = \sqrt{P}\boldsymbol{h}\boldsymbol{w}_r x + z, \quad (6)$$

where $P$ is the transmit power at the BS, $x$ is the transmitted signal with a unit power, $\boldsymbol{w}_r \in \mathcal{C}^{N_T \times 1}$ is the beamformer of the BS-RIS link and has unit power (as an upper bound, while in practice, this may not be achievable), and $z$ is the additive Gaussian noise at the receiver side. Also, ignoring the direct BS-UE link as well as the nonlinearities between adjacent reflectors, the equivalent channel $\boldsymbol{h} \in \mathcal{C}^{1 \times N_b}$ in the RIS-assisted link (BS-RIS-UE) is given by

$$\boldsymbol{h}_r = \boldsymbol{h}_{ru}\boldsymbol{\Theta}\boldsymbol{h}_{br}. \quad (7)$$

Here, $\boldsymbol{h}_{br} \in \mathcal{C}^{M \times N_b}$ and $\boldsymbol{h}_{ru} \in \mathcal{C}^{1 \times M}$ are the channel between BS-RIS and RIS-UE, respectively. Moreover,

$$\boldsymbol{\Theta} = \text{diag}\left(e^{j\theta_1}, \ldots, e^{j\theta_M}\right) \quad (8)$$

is the reflection coefficient matrix of the RIS. With joint BS active and RIS passive beamforming, the data rate at UE with RIS assistance can be expressed as

$$R = Q\log_2\left(1 + \frac{P|\boldsymbol{h}_{ru}\boldsymbol{\Theta}\boldsymbol{h}_{br}\boldsymbol{w}_r|^2}{\sigma^2}\right). \quad (9)$$

One of the main motivations of using RIS, and not small access points such as relays, NCRs, and IABs, is the cost and energy reduction. With a cheap node, however, hardware imperfections may affect the reflection quality of the RISs. Particularly, it is likely that, in practice, RISs may provide an imperfect reflection because of different HWIs, such as non-linear amplifier, phase error, quantization noise, where the performance of the RIS may be degraded [21]-[22]. Also, the non-linearities introduced from an RIS with discrete reflection elements will pollute not only the own network but other networks as well. Here, we use the HWI model from [23, Eq. (10)], where the data rate (9) can be rephrased as

$$R_{\text{HWI}} = Q\log_2\left\{1 + \frac{\beta M^2 + \xi M + \mu_{bu}}{(\kappa_t + \kappa_r)(\beta M^2 + \xi M + \mu_{bu}) + \frac{\sigma^2}{P}}\right\}, \quad (10)$$

in the cases with HWI. Here, $\beta = 4\alpha^2 \mu_{ru}\mu_{br}/\pi^2$, and

$$\xi = \left(1 - \frac{4}{\pi^2}\right)\alpha^2 \mu_{ru}\mu_{br} + \frac{4\alpha}{\pi}\mu_{br}^{\frac{1}{2}}\mu_{ru}^{\frac{1}{2}}\mu_{bu}^{\frac{1}{2}}\cos(\varphi_{bu}). \quad (11)$$

**Algorithm 1** Joint beam optimization using AO in RIS systems

**Requires**: $h_{br}$ and $h_{ru}$
1. Initialize $w_r$ to some feasible values at the BS.
   **Repeat**
2. Calculate $h_w = h_{br} w_r$;
3. Set $\phi_m = -\angle\{h_{ru}^*(m) h_w\}$;
4. Calculate $h_{ru} \Theta h_w$
5. Set $w_r$ as the right dominant eigenvector of $h_{ru} \Theta h_w$
   **Until** $R$ in (9) Convergence

**Return** Optimal rate $R$ for UE (9)

---

**Algorithm 2** Joint beam optimization using DFT codebook-based beamforming in RIS systems

**Requires**: Concatenate channel $h_{ru} \Theta h_{br} w_r$ with selected $\Theta$, and a pre-defined DFT codebook $V \in \mathcal{C}^{M \times M}$

**For** $i = 1: M$ **do**
1. The RIS uses $i$-th beam $v_i$ from the pre-defined codebook $V$ and forms the reflection matrix as $\Theta = \text{diag}(v_i)$;
2. The UE calculates the received power for the selected RIS beam $\|h_{ru}\text{diag}(v_i) h_{br}\|^2$;
   **End For**
3. The UE feeds back the best beam index $i_{best}$ to the BS;
4. The BS obtains the precoder as $w_{DFT} = h_{DFT}^H / \|h_{DFT}\|$, where $h_{DFT} = \|h_{ru}\text{diag}(v_{i_{best}}) h_{br}\|^2$;

**Return** Optimal rate $R$ for the UE (9)

---

Here, $\alpha \in (0,1]$ is the RIS amplitude coefficient which describes the qualification of the RIS reflection. Also, $\mu_{BR}$, $\mu_{RU}$, and $\mu_{BU}$ are the power attenuation coefficient of the BS-RIS, the RIS-UE, and the BS-UE link (with negligible NLoS link), respectively, and they can be represented by the path loss factor of each channel. Moreover, $\varphi_{BU}$ is the phase shift of the channel $h_{BU}$. Finally, $\kappa_t$ and $\kappa_r$ are the parameters describing the severity of the distortion noises at the transmitter and receiver, respectively (see [23] for details).

For the RIS-aided indirect link, we need to jointly optimize the active beamforming vector $w_r$ at the BS, as well as the passive beamforming matrix $\Theta$ at the RIS, according to

$$\max_{w_R, \Theta} R$$
$$\text{s.t. } \|w_r\|^2 = 1, \quad (12)$$
$$\theta_m \in [0, 2\pi], m = 1, \ldots, M, \forall m.$$

For the single link evaluation, we consider two different RIS optimization schemes, namely, alternating optimization (AO) and discrete Fourier Transform (DFT) codebook-based optimization. The problem (13) has been widely studied in the literature under different setups. For instance, with perfect knowledge of CSI, i.e., $h_{br}$ and $h_{ru}$, AO has shown to approach optimal solution of (13) [24, Sec. III-B], [25, Sec. III], [26, Algorithm I]. Note that, imperfect CSI is indeed a more practical assumption and would affect the performance. In this paper, we only consider the cases with perfect and no CSI in Algorithms 1 and 2, respectively.

As shown in Algorithm 1, maximizing $R$ is equivalent to optimizing $|h_{ru} \Theta h_{br} w_r|$. Using AO, for a fixed $w_r$, the optimal $\theta_m$ results in $-\angle\{h_{ru}^*(m) h_w\}$ with $h_w = h_{br} w_r$. Then, for fixed $\Theta$, the optimal $w_r$ can be obtained by calculating the dominant right eigenvector of $h_{ru} \Theta h_w$. As an alternative method and inspired by the precoding scheme designed in [27, Algorithm 1], one can use a DFT codebook-based beam optimization where the RIS beam is selected from the pre-defined beam patterns while only the active beamforming $w_r$ is needed to be optimized, as presented in Algorithm 2. Here, Algorithm 2 does not require explicit CSI when optimizing the beams.

## IV. PERFORMANCE EVALUATION

We divide the simulation results into two parts. In Figs. 3-5, we study the link-level performance of the NCR/RIS-assisted network with the setup given in Fig. 2. Then, in Fig. 7, we study the effect of interference in an NCR-assisted network with the setup given in Fig. 6. We consider a generic mmw channel model with angel-of-departure (AoD) $\psi$, angle of arrival (AoA) $\phi$ and multipath

$$h = \sqrt{\frac{PL}{L}}\sum_{l=1}^{L} \iota_l \boldsymbol{a}(\psi_l) a^T(\phi_l). \quad (13)$$

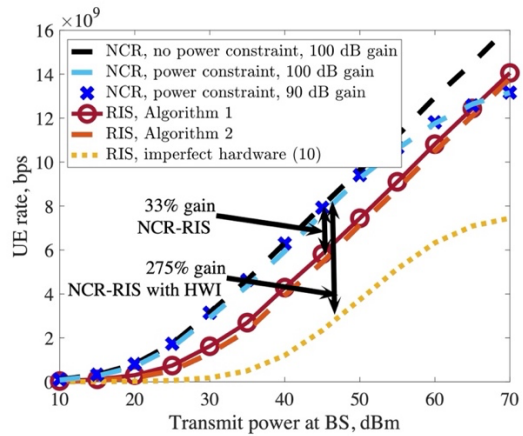

Fig. 3. The UE data rate vs the BS transmit power. $\kappa_t = \kappa_r = 0.05^2$, $\alpha = 1$, $\varphi_{bu} = \pi/4$. The positions of the BS, RIS/NCR, and the UE are set to [140 m, 50 m], [0, 0], and [100 m, 0], respectively.

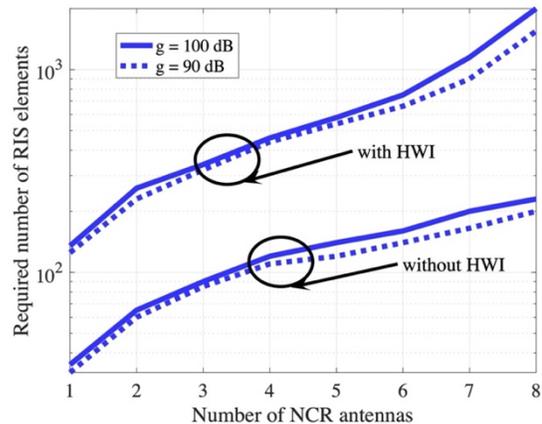

Fig. 4. The required number of RIS elements to achieve the same performance as in the NCR-assisted network. The maximum output power constraint at the BS and the NCR are set to 43 and 40 dBm, respectively. $\kappa_t = \kappa_r = 0.05^2$, $\alpha = 1, \varphi_{bu} = \pi/4$.

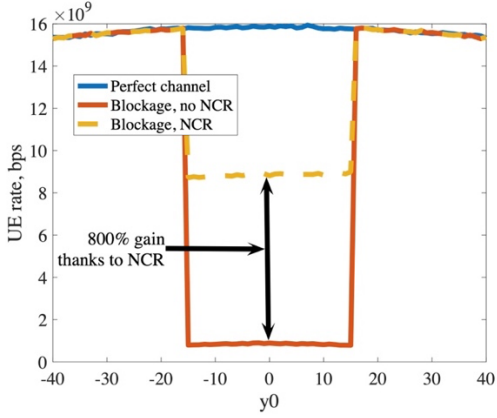

Fig. 5. An illustration of how NCR helps the network in the presence of blockage. Considering Fig. 2, one BS is located at [0, 0] and one NCR is located at [40 m, 20 m]. The user moves in parallel with y-axis with [60 m, $y_0$ m]. One blocker is located from [40 m, 40 m] to [-10 m, 10 m].

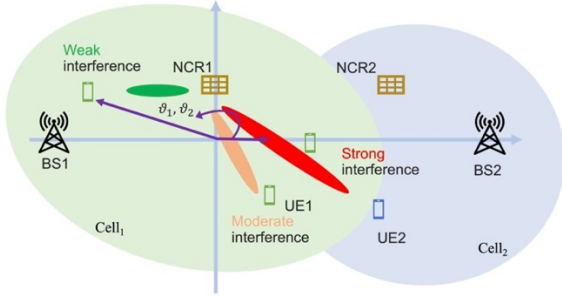

Fig. 6. Simulation setup for two cells.

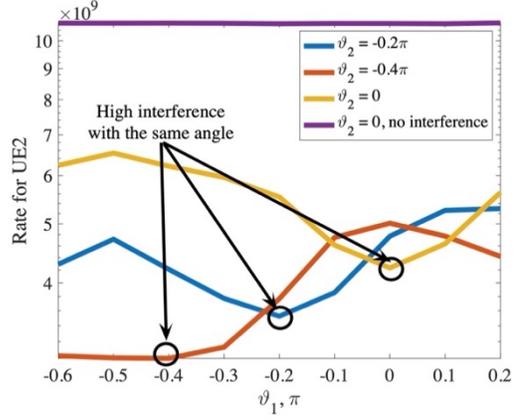

Fig. 7. The throughput of $UE_2$ with different angles of $UE_1$ in Fig. 6. Here, the BS transmit power $P$ = 43 dBm, $N_{NCR}$ = 10, $g$ = 100 dB. The AoDs w.r.t. [0, 0] to $UE_1$ and $UE_2$ are represented by $\vartheta_1$ and $\vartheta_2$, respectively. Also, the positions of $UE_2$ are [160 m, -40 m], [160 m, -20 m] and [160 m, 20 m] for the cases with $\vartheta_2 = -0.4\pi, -0.2\pi$, and 0, respectively.

Here, $\iota \sim \mathcal{CN}(0,1)$ and $L$ is the number of paths. Also, $a(\psi_l) = [1, \ldots, e^{jkd_a(n_t-1)\sin(\psi_l)}]$ $a(\phi_l) = [1, \ldots, e^{jkd_a(n_r-1)\sin(\psi_l)}]$ are the antenna steering vectors with $k = 2\pi/\lambda$, $d_a = \lambda/2$, where $\lambda$ is the wavelength, and $n_t/n_r$ are the transmit/receive antennas, respectively. The path loss at distance $d$ can be obtained by the mmMAGIC model [28, Table IV] for the LoS case

$$PL_{LoS} = 19.2 \log_{10}(d) + 32.9 + 20.8 \log_{10}(f_c), \quad (14)$$

and for NLoS

$$PL_{NLoS} = 45 \log_{10}(d) + 31 + 20 \log_{10}(f_c). \quad (15)$$

The number of paths $L$ is set to 3 in the simulations. Moreover, the codebook-based beamforming proposed in Algorithm 2 can be applied with different codebooks. Here, we present results with [29]

$$V = \{\omega_i\}_{i=1}^{\sqrt{N}} \otimes \{\nu_j\}_{j=1}^{\sqrt{N}}, \quad (16)$$

where

$$\omega_i = \left[1, e^{\frac{2\pi(i-1)}{\sqrt{N}}}, \ldots, e^{\frac{2\pi(i-1)(\sqrt{N}-1)}{\sqrt{N}}}\right], \quad (17)$$

$$\nu_j = \left[1, e^{\frac{2\pi(j-1)}{\sqrt{N}}}, \ldots, e^{\frac{2\pi(j-1)(\sqrt{N}-1)}{\sqrt{N}}}\right], \quad (18)$$

and $\otimes$ represents the Kronecker product.

In Figs. 3-5, we consider one BS with $N_b$ = 16 antennas, one RIS with $M$ = 100 elements/one NCR with 8 antennas at each side. Unless otherwise stated, the maximum output power constraint at the NCR is set to 40 dBm while the amplification gain is 90 or 100 dB (Note that in Rel.17 RF repeaters, the amplification gain is considered to be 90 dB. While not yet specified in 3GPP, the same or a different amplification gain may be considered for NCRs). The carrier frequency $f_c$ is set to 28 GHz with an ideal 1 GHz channel bandwidth $Q$. The noise power is set as -174 dBm/Hz with 10 dB noise figure. The antenna gains of the BS, RIS/NCR, and the users are set to 18 dBi, 18 dBi, and 0 dBi, respectively. For the HWI-affected rate (10), in simulation we set $\kappa_t = \kappa_r = 0.05^2$, $\alpha = 1$, $\varphi_{bu} = \pi/4$ in harmony with [23].

Fig. 3 compares the performance of the RIS- and NCR-assisted setup of Fig. 2 in terms of the UE rate. As shown, the proposed AO- and DFT-based RIS optimization methods can reach almost similar performance for a broad range of parameters, while the DFT method does not require explicit CSI when optimizing the beams. The effect of NCR maximum output constraint is observed only at very high BS transmit powers. Without a maximum output power constraint at the

NCR, NCR-assisted network achieves higher throughput compared to the RIS-assisted network. For instance, with 45 dBm BS transmit power, the implementation of the NCR leads to 33% rate improvement, compared to the cases with ideal RISs. With a maximum output power constraint, however, the RIS-assisted setup outperforms the NCR-assisted setup only at very high BS transmit powers, if the no HWI is considered for the RIS. On the other hand, HWI deteriorates the performance of the RIS-assisted setup significantly (Fig. 3). For instance, with the parameter setting of the figure and 45 dBm BS transmit power, the implementation of the NCR improves the achievable rate of the HWI-affected RIS setup by 275%.

Fig. 4 shows the required number of RIS elements to achieve the same throughput as in the NCR-assisted setup. Here, the results are presented for both cases with ideal and HWI-affected RIS models. Also, the transmit power at the BS and the maximum output power constraint at NCR are set to 43 dBm and 40 dBm, respectively with 90 dB/100 dB NCR amplification gain. As demonstrated, to achieve the same performance as in the NCR-assisted setup, a large number of RIS elements are required, and the required numbers increase with the NCR amplification gain. For instance, compared to the cases with 7 antennas at the NCR with amplification gain of 100 dB, around 200 RIS elements are required to obtain equivalent performance as in the NCR-assisted setup, if the RIS is ideal. With HWI, however, the required number of RIS elements increases significantly, for instance, to more than 1000 RIS elements in the considered example.

Fig. 5 gives an illustration of how the NCR helps the network in the presence of blockage assuming no additional BSs are helping to bypass the blockage (In practice, the neighbor BSs can also help in bypassing blockages [30]). Here, considering Fig. 2, one BS is located at [0, 0] and one NCR is located at [40 m, 20 m]. The UE moves in parallel with the $y$-axis with [60 m, $y_0$ m]. One blocker is a wall located from [40 m, 40 m] to [-10 m, 10 m]. As demonstrated in Fig. 5, with the mmw band, the blockage affects the UE achievable rate significantly. However, the presence of the NCR helps to bypass the blockage and avoid large performance drop. For instance, with the considered parameter settings of Fig. 5, the presence of the NCR improves the UE achievable rate by 800%, compared to the cases with no NCR.

In Fig. 7, we study the effect of the interference in NCR-assisted networks. Particularly, as shown in Fig. 6, we focus on the case where at the cell edge, $UE_2$ at $Cell_2$ is affected by non-intended signal of $Cell_1$ forwarded by $NCR_1$. In this way, the $UE_2$ throughput calculation should consider signal-to-interference-plus-noise ratio (SINR) with interference from $NCR_1$ instead of SNR in (5). Here, considering Fig. 6, $BS_1$, $BS_2$, $NCR_1$, and $NCR_2$ are located at [-50 m, 0], [250 m, 0], [0, 20 m], and [200 m, 20 m], respectively. Also, the UE2 rate is evaluated in different positions of UE2 in [160 m, -40 m], [160 m, -20 m] and [160 m, 20 m] for the cases with AoDs being around -0.4π, -0.2π, and 0, respectively. $UE_1$ moves in a circle of radius 80 m for a broad range of angles.

As seen in Fig. 7, while $UE_1$ benefits from the presence of the NCR, the cell edge $UE_2$ may be affected by the interference signal forwarded by $NCR_1$, depending on the position of the UEs. This may end up in considerable rate drop for UE2. For this reason, it may be beneficial to develop interference management mechanisms in NCR-assisted networks.

## V. OPEN RESEARCH PROBLEMS

In general, amplify-and-forward relaying is not a new topic, e.g., [31]-[32], and different works have studied the effect of beamforming on performance of multi-antenna AF relay networks, e.g., [33]-[35]. For this reason, we mainly concentrate on the industrial open problems to be discussed in 3GPP Release 18 work-item. Particularly, from RAN1/RAN2 perspective, the NCR work-item concentrates on specifying the signaling and behavior of the following side control information for controlling the NCR-Fwd [17]:

- Beamforming: Here, beamforming considers both the backhaul and the access links of the NCR while, compared to the access link, beamforming in the backhaul link, i.e., between the gNB and the NCR-Fwd, may be simpler and less dynamic.

- UL-DL TDD configuration: As agreed in 3GPP RAN1 meeting 109e [19], at least semi-static TDD UL/DL configuration is needed for the NCR for links including C-link, backhaul link and access link. Also, the same TDD UL/DL configuration is always assumed for backhaul link and access link. The same TDD UL/DL configuration is assumed for C-link and backhaul link and access link if NCR-MT and NCR-Fwd are in the same frequency band. However, how to handle the flexible symbols may require further discussions in 3GPP.

- ON-OFF information: 3GPP has agreed that ON-OFF information is beneficial and recommended for the NCR to control the behavior of NCR-Fwd. Here, ON-OFF configuration may be beneficial for efficient interference management or improved energy efficiency. However, the detailed mechanism of ON-OFF indication and determination is still under discussions.

Indeed, 3GPP will specify the required signaling which will enable the features of the considered objectives. Also, while the main focus of the work-item is on RAN1 aspects of the NCRs, 3GPP will specify the solution of NCR management (i.e., the identification and authorization/validation of NCR) from the RAN2/RAN3 perspectives. Finally, RAN2 and RAN4 may study the radio resource management (RRM) functions to be supported and specify the RRM requirements of NCR-MT, if necessary, and RAN4 may study and specify the RF and EMC requirements of NCR, if necessary. Here, cost efficiency is a key consideration point for the NCRs, where it is aimed to keep the complexity limited.

Finally, while there are many theoretical works on AF relays and it is not required in 3GPP Rel. 18, still system-level simulations may be beneficial to evaluate the potentials and challenges of NCRs, e.g., in comparison with other technologies such as IAB or RF repeaters.

## VI. Conclusions

We introduced the concept of NCRs, which is considered in 3GPP Release 18. Particularly, introducing the terminologies and the objectives considered by 3GPP, we presented conceptual simulations and compared the performance of the NCR-assisted networks with the cases using RISs. As we discussed, there are still different open problems to be addressed in NCRs.

## Acknowledgment

This work was supported by the European Commission through the H2020 Project Hexa-X under Grant 101015956, and in part by the Gigahertz-ChaseOn Bridge Center at Chalmers in a project financed by Chalmers, Ericsson, and Qamcom.